\documentclass[twocolumn,showpacs,preprintnumbers,amsmath,amssymb,prb]{revtex4}
\usepackage{graphicx}
\usepackage{dcolumn}
\usepackage{bm}
\newcommand{\CuTeX} {{\mbox{Cu${}_2$Te${}_2$O${}_5$X${}_2$}}}
\newcommand{\CuTeCl} {{\mbox{Cu${}_2$Te${}_2$O${}_5$Cl${}_2$}}}
\newcommand{\CuTeBr} {{\mbox{Cu${}_2$Te${}_2$O${}_5$Br${}_2$}}}
\newcommand{\half}{{\ensuremath{\frac{1}{2}}}}
\newcommand{\trhalf}{{\ensuremath{\frac{3}{2}}}}
\renewcommand{\deg}{\ensuremath{^{\circ}}}
\newcommand{\Ps}{\mbox{\bf P}} 
\newcommand{\Pin}{\mbox{\bf P$'$}} 
\newcommand{\Q}{\ensuremath{\mathbf{M}_{\perp}}} 
\newcommand{\Qs}{\ensuremath{\mathbf{M}_{\perp}^{*}}}
\begin{document}
\preprint{APS/123-QED}

\title{Incommensurate magnetic ordering in ${\CuTeX}$ (X=Cl, Br)
studied by single crystal neutron diffraction}
\author{O. Zaharko,H. R\o{}nnow, J. Mesot}
\affiliation{Laboratory for Neutron Scattering, ETHZ \& PSI, CH-5232
Villigen, Switzerland}
\author{S. J. Crowe, D. M$^c$K. Paul}
\affiliation{Department of Physics, University of Warwick, Coventry, CV4 7AL,
United Kingdom}
\author{P. J. Brown}
\affiliation{Institut Laue-Langevin, 156X, 38042 Grenoble C\'{e}dex, France}
\author{A. Daoud-Aladine}
\affiliation{ISIS Facility, Rutherford Appleton
Laboratory, Chilton, Didcot,  Oxfordshire OX10 OQX, United Kingdom}
\author{A. Meents, A. Wagner}
\affiliation{Paul Scherrer Institute, 
CH-5232 Villigen, Switzerland}
\author{M. Prester}
\affiliation{Institute of Physics, P.O.B.304, HR-10 000, Zagreb, Croatia}
\author{H. Berger}
\affiliation{Institut de Physique de la Mati\`{e}re Complexe,
EPFL,CH-1015 Lausanne, Switzerland}
\date{\today}

\begin{abstract}
Polarized and unpolarized neutron diffraction studies have been
carried out on single crystals of the coupled spin tetrahedra systems
${\CuTeX}$ (X=Cl, Br). A model of the magnetic structure associated
with the propagation vectors
${\bf{k'}_{Cl}}\approx(-0.150,0.422,\half)$
and ${\bf{k'}_{Br}}\approx(-0.172,0.356,\half)$ and stable below
$T_{N}$=18 K for X=Cl and $T_{N}$=11 K for
X=Br is proposed. A feature of the model, common to both the bromide
and chloride, is a canted coplanar motif for the  4 Cu$^{2+}$
spins on each tetrahedron which rotates on a helix from cell to cell
following the propagation vector.
The Cu${}^{2+}$
magnetic moment determined for  X=Br, 0.395(5)$\mu_B$,
is significantly less than for X=Cl,  0.88(1)$\mu_B$  at 2~K. The
magnetic structure of the chloride associated with the wave-vector
${\bf{k'}}$ differs from that determined previously for the wave vector 
${\bf k}\approx(0.150,0.422,\half)$ [O. Zaharko, et.al., Phys. Rev. Lett. $\bf{93}$, 217206 (2004)].

\end{abstract}

\maketitle

\section{\label{sec:level1}Introduction}
Systems with weakly interacting frustrated magnetic clusters form an
interesting class of materials with properties lying between those of
quantum spin systems and classical magnets.\cite{Diep03} In this context
the ${\CuTeX}$ (X=Cl, Br)\cite{Johnsson00} compounds have recently
attracted strong interest, as they contain Cu${}^{2+}$ tetrahedral
clusters. The antiferromagnetic exchange interactions between the spins
within a tetrahedron are geometrically frustrated and the coupling
between the tetrahedra was assumed to be weak. Whilst the excitation
spectrum of isolated tetrahedra is well known to be gapped, the presence
of even small anisotropy or inter-tetrahedral coupling may lead to
interesting new ground states and excitations. In these compounds the
magnetic susceptibility reaches a maxima at $T\sim$ 25 K before dropping
sharply at low temperatures, which was attributed to the presence of a
singlet-triplet spin-gap.\cite{Johnsson00,lemmens} Further evidence of
spin-gapped behavior in the bromide is observed in Raman
scattering,\cite{lemmens,gros,jensen} which also reveals evidence of
what is suggested to be a low energy longitudinal magnon. Fitting the
susceptibility data to an isolated tetrahedral model with four nearest
neighbour ($J_{1}$) and two next nearest neighbour ($J_{2}$) exchange
interactions gives the coupling strengths $J_1$=$J_2\sim$ 43 K  and 38.5
K for X = Br and Cl respectively. However,
susceptibility,\cite{Johnsson00, lemmens} heat
capacity\cite{lemmens,gros} and thermal
conductivity\cite{prester,sologubenko} measurements all show evidence of
magnetic ordering at low temperatures, $T_{N}$=18 K (Cl) and 11 K (Br),
which requires inter-tetrahedral couplings. The effect of
inter-tetrahedral coupling and the relative strengths of exchange
interactions in this system have been investigated theoretically by band
structure calculations,\cite{valenti} spin dimer analysis\cite{whangbo},
flow-equation method\cite{brenig}  and a mean field
analysis.\cite{gros,jensen} The consequences of antisymmetric
Dzyaloshinsky-Moriya (DM) interactions have been also analysed.\cite{kotov}
Experimentally, magnetic excitations with a dispersive component are
observed in both compounds by inelastic neutron scattering
measurements,\cite{Crowe,Zaharko05} which are associated with the
development of long range order. \\
The ground state magnetic structure
is found from neutron diffraction studies to be rather
complex.\cite{Zaharko04} Both compounds have incommensurate magnetic
structures with wave-vectors ${\bf{k'}_{Cl}}\approx(-0.150,0.422,\half)$
and ${\bf{k'}_{Br}}\approx(-0.172,0.356,\half)$. For ${\CuTeCl}$ the
co-existence of two different magnetic structures with
${\bf{k'}}=(-k_x,k_y,\half)$ and  ${\bf{k}}=(k_x,k_y,\half)$ has been
detected.
In the model proposed for the $\bf{k}$ structure the four
Cu$^{2+}$ ions of each
tetrahedron form two pairs with the spins on the two ions of a pair rotating in
the same plane with a constant canting angle between them. The canting angles
were determined as
38(6)\deg\ for the first and 111(10)\deg\ for the second pair.\\

\section{New diffraction results}

\begin{figure}[tbh]
\caption {The $hk0$ and $hk\half$ layers of reciprocal space of a
$\CuTeX$ crystal. Black circles correspond to magnetic reflections
$\{hk\frac12\}$ of ${\bf{k'}}$ and grey circles - of ${\bf{k}}$ wave
vectors. The circle radii are proportional to intensities of magnetic
reflections. Dashed circles with center at 000 connect reflections with
intensity $I_1$ and $I_2$.}
\includegraphics[width=86mm,keepaspectratio=true]{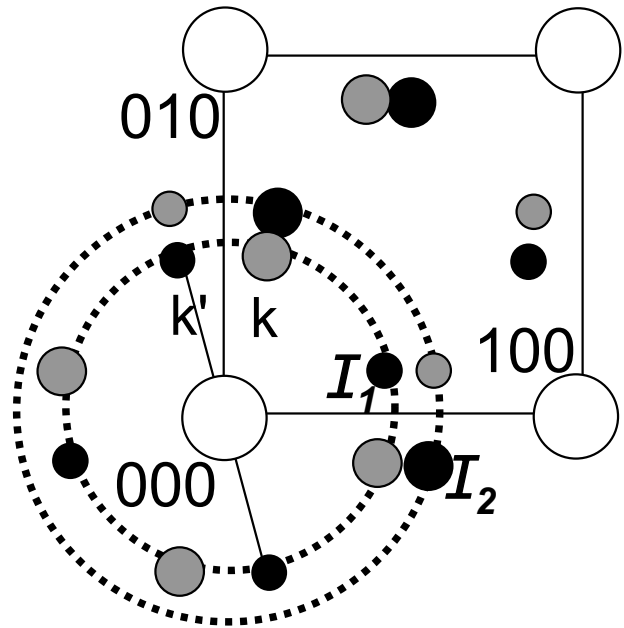}
\label{fig1}
\end{figure}

To obtain a more complete picture of the ground states of ${\CuTeX}$ we
have made several new diffraction experiments.
An X-ray diffraction experiment
was carried out on a ${\CuTeCl}$ single crystal ($\sim$10
$\mu$m${{}^3}$) at 10 K and 25 K at the X10
beamline ($\lambda=0.71073$ \AA) at the SLS synchrotron. It revealed that the
crystal structure of ${\CuTeCl}$ at temperatures below and above $T_{N}$, has
the same tetragonal space group $P\overline{4}$, as it does at 300
K.\cite{Johnsson00} The group is  non-centrosymmetric and racemic twins were
found to be present with the volume ratio 37(4):63(4). No features which could
explain the co-existence of wave vectors ${\bf{k}}$ and ${\bf{k'}}$ below
$T_{N}$ were detected. The possibility of growth twins related by 
reflection in {100} planes was considered. No evidence for such 
twinning was obtained in the structure refinements confirming that the
$\{$hkl$\}$ and $\{$-khl$\}$ families of reflections are independent.
\\
Spherical
neutron polarimetric measurements were made at 2 K on a ${\CuTeCl}$ single
crystal (6 x 2.8 x 2.6
mm${{}^3}$) with CRYOPAD II installed on the D3 diffractometer at ILL
($\lambda=0.843$ \AA).
These were supplemented by unpolarized
integrated intensity measurements at 2 K on the same ${\CuTeCl}$ crystal and on
a ${\CuTeBr}$ (4 x 1 x 1 mm${{}^3}$) crystal using the D10
diffractometer at ILL ($\lambda=2.359$ \AA).\\
The magnetic diffraction patterns given by the various ${\CuTeX}$ crystals
studied so far show an important qualitative difference.
The ${\CuTeCl}$ crystal used in the present experiment gave 4 magnetic
reflections at the lowest  $2\theta$ value (black spots in Fig.~\ref{fig1}).
They originate
from two configuration domains with wave vectors
${\bf{k'}}$ related by the four-fold axis (rotation of 90\deg). For the
${\CuTeBr}$ crystal on the other hand only 2 reflections, corresponding to a
single configuration domain, were found at the lowest angle  (black spots
connected by solid lines in Fig.~\ref{fig1}). It should be recalled
that in the previous study\cite{Zaharko04} 8 reflections from two
configuration domains of two independent wave vectors ${\bf{k'}}$ and
${\bf{k}}$
were reported (black and grey spots in Fig.~\ref{fig1}).\\
There is a quantitative difference between
the intensities of corresponding reflections obtained from the propagation
vectors
${\bf{k}}$ and ${\bf{k'}}$. This difference can be clearly seen by comparing
the intensities $I_{1}$ and $I_{2}$ of the reflections $000+{\bf{k}}$ and
$100-{\bf{k}}$. For the  ${\CuTeCl}$ and ${\CuTeBr}$ crystals studied in the
present experiment
the inequality $I_{1}<I_{2}$ holds. In the previous study of another
${\CuTeCl}$
crystal
\cite{Zaharko04} this relation was the same $I_{1}<I_{2}$ for the
${\bf{k'}}$ set, but opposite $I_{1}>I_{2}$ for the ${\bf{k}}$ set; and it was
from this latter set that the ${\bf{k}}$ structure was derived.\cite{remark1}\\
In what follows we will describe the experimental
observations made on wave vector ${\bf{k'}}$ and will use them to derive a
model
for the ${\bf{k'}}$ magnetic structure.
\\
For polarimetric measurements the crystal was mounted
with the $a^*$-direction vertical inside an ILL orange cryostat and
cooled to 2 K. As the wave vector
${\bf{k'}}\approx[-0.150,0.422,\half]$ has a small $a^*$ component,
the beam scattered by a magnetic reflection $hkl$ with $h=0.15$ is
tilted from the horizontal plane by an angle $\nu\approx 1$\deg\ and can be
measured by ensuring that the vertical aperture of the detector is wide
enough.  The inclination $\rho$ of the scattering vectors to the
horizontal plane is $\rho$=11.6\deg\ for ${k=0.422,l=\half}$ and
$\rho$=4.6\deg\ for ${k=0.422,l=\trhalf}$.
The polarization of the scattered
beam was measured using
a spin polarized ${}^{3}$He filter. The filter polarization decayed with
a time constant of $\approx$100 h and its effective polarization
transmission varied from $\sim$0.73 to $\sim$0.55 between filter
changes. The decay was followed by measuring the polarization scattered
by the 002 nuclear reflection at regular intervals and an appropriate
correction was applied to the scattered polarizations. \\
Measurements of each reflection were made with the incident neutrons
polarized successively in three directions: parallel to the vertical $z$
direction ($P'_{z}$), along the $x$ horizontal component of the
scattering vector $\bf{q}$ ($P'_{x}$) and in the $y$ direction that
completes the right-handed Cartesian set. In this polarization
coordinate system the magnetic interaction vector
(${\bf{M_{\perp}}}=\bf{C}+i\bf{D}$), which is the  projection of the
Fourier transform of the magnetization ${\bf{M(r)}}$ onto the plane
perpendicular to $\bf{q}$, lies mainly in the $yz$-plane. For each
incident polarization direction the components of scattered polarization
parallel to the $x, y$ and $z$ directions were determined.
The six magnetic reflections labelled $\bf{h_{1}}$-$\bf{h_{6}}$
in Table~\ref{tab1} were studied.
Their intensities were
rather weak, and the background at low $2\theta$ rather high,
especially for the reflections $\bf{h_{1},h_{2},h_{3}}$, so reasonable
statistics
could only be obtained by measuring for about 6 hours/reflection.
Even a qualitative analysis of the scattered polarization sets a number of
valuable constraints on possible models for the magnetic structure of
${\CuTeCl}$.\\
The polarization $\Ps$ of neutrons scattered by a pure magnetic
reflection
can be written as:\cite{Blume, Maleev}
\begin{eqnarray}
\Ps I
&=&-\Pin(\Q\cdot\Qs)+2\Re(\Q(\Pin\cdot\Qs))
\nonumber\\
&&-2\Im(\Q\times\Qs)
\label{eq:eq1}
\end{eqnarray}
with the first term parallel to $\Pin$, the second to $\Q$ and the
third to ${\bf{q}}$.
$I$ is the scattered intensity which contains a polarization
independent and a polarization dependent term:
\begin{equation}
{I = \Q\cdot\Qs +
\Pin\cdot\Im(\Q\times\Qs)}
\label{eq:eq2}
\end{equation}\\
These equations rewritten in the form of polarization
matrices:\cite{Brown01}

\begin{equation}P_{ij}=\left|\begin{array}{ccc}
-\frac{M^2+J_{yz}}{M^2+J_{yz}}&0&0\\ -\frac{J_{yz}}{M^2+J_{yz}}&
\frac{-M^2+R_{yy}}{M^2}& \frac{R_{yz}}{M^2}\\
-\frac{J_{yz}}{M^2+J_{yz}}& \frac{R_{yz}}{M^2}&
\frac{-M^2+R_{zz}}{M^2} 
\end{array}
\right|\label{eq:eq3}
\end{equation}
with $M^2={\bf{M_{\perp}}\cdot \bf{M^*_{\perp}}}$,
$R_{ij}=2\mathfrak{R}(M_{\perp i} M^*_{\perp j})$ and
$J_{ij}=2\mathfrak{J}(M_{\perp i} M^*_{\perp j})$ can be directly
compared with the results presented in Table~\ref{tab1}. \\
Before attempting a detailed analysis, the different types of magnetic
domains which can be present  in ${\CuTeX}$ should be considered. It is worth
remembering\cite{Zaharko05} that configuration, orientation and chiral
magnetic domains are all possible. The configuration domains
give rise to separate sets of magnetic peaks and are not important
for the polarization data analysis. Orientation domains can occur if
the $2_z$-axis is not in the
magnetic symmetry group. Chirality domains are present because the
propagation vector $\bf{k'}$ is not one half of a reciprocal lattice
vector  (${\bf k' }\ne {\bf g-k'}$). They are not the same as racemic twins
which are allowed because the  inversion center is missing in
the crystallographic
space group. The last two types of domain contribute to the same
magnetic peaks and their presence could significantly complicate the
D3 data analysis since
orientation domains can depolarize the scattered beam and chirality
domains can conceal the special features of helical structures.\\
It was found that within the statistical accuracy  the scattered beam was
fully polarized for all measured reflections which suggests that only
one orientation domain is present. The two chiral domains are also
unequally populated, since the $\bf{h_{1}}$ and $\bf{h_{2}}$ reflections were
almost absent for incident polarization $P'_{+x}$, but
had significant intensity for $P'_{-x}$. \\
Furthermore, based on the polarization
data we can immediately deduce the type of magnetic structure. The very
presence of the $x$-components in the scattered polarization ($P_{xy}$
and $P_{xz}$) indicates rotation of polarization towards $\bf{q}$, which
is not compatible with any amplitude modulated or collinear  structure.
Such structures would have ${\bf{M_{\perp}} \parallel
\bf{M^{*}}_{\perp}}$ and the neutron polarization would only precess by
180\deg\ about ${\bf{M_{\perp}}}$. Therefore the magnetic structure must be
helical.\\
The most significant qualitative conclusions from the polarimetric
measurements are as follows:
\begin{itemize}
\item[1.]
{The scattered polarization for the reflection $\bf{h_{2}}$ with
$\bf{q_{2}} {\parallel\bf{k'}}$, shows that the structure is not
composed of helices with spins rotating in a plane normal to the wave
vector ${\bf{k'}}$, for in such a case all polarization would be
rotated towards the $x$ direction.
The presence of  $y$ and $z$ components of scattered polarization
indicates that one or more of the planes in which the spins rotate (plane of
helices) must be inclined to
the wave vector.}
\item[2.]
{For all the reflections studied the $P_{zz}$ components
are positive, while $P_{yy}$ are negative. i)
This clearly indicates that all ${\bf{M_{\perp}}}$ vectors have a
$z$-component,
so there must be a component of the magnetic moment along $a$.
ii) It also means that the $z$-components of ${\bf{M_{\perp}}}$ are larger than
the $y$-components ($C^2_{z}+D^2_{z} > C^2_{y}+D^2_{y}$). This
might indicate that the planes of the  helices are close to or contain
the $a$-axis.}
\item[3.]
{The magnitude of the $P_{yy}$ and $P_{zz}$ components tend to
be larger for reflections with $l=\trhalf$ than for those with
$l=\half$ which strongly suggests the existence of a $c$-component of
the magnetic moment.}
\end{itemize}
\begin{table}
\caption{
Polarization matrices $P_{ij}$ measured for ${\bf{k'}}$ magnetic reflections of
${\CuTeCl}$ at 2 K ($i$ - incoming,  $j$ - outcoming component of
polarization). I is measured intensity.
\label{tab1}}
\begin{ruledtabular}
\begin{tabular}{cccccccc}
\multicolumn{3}{c}{${h~~~~k~~~~l}$}&${P'_{i}}$&${P_{ix}}$&${P_{iy}}$&${P_{iz}}$&
I\\
\\
0.15&-0.42&${\half}$&-$x$&0.93(5)&-0.04(5)&0.19(4)&5.0(3)\\
$\bf{h_{1}}$&&&$y$&0.83(9)&-0.71(9)&0.03(8)&3.0(2)\\
&&&$z$&0.6(1)&-0.2(1)&1.0(1)&2.2(2)\\
&&&-$z$&1.2(1)&0.18(8)&-0.34(10)&2.2(2)\\
\\0.15&-0.42&-${\half}$&-$x$&0.93(8)&-0.03(9)&0.33(7)&4.5(2)\\
$\bf{h_{2}}$&&&$y$&0.8(2)&-0.5(2)&0.6(2)&2.7(1)\\
&&&$z$&0.7(2)&-0.0(2)&1.1(2)&2.3(1)\\
&&&-$z$&0.7(1)&-0.2(1)&-0.3(2)&3.1(4)\\
\\-0.15&-0.58&${\half}$&$x$&-1.03(3)&-0.04(3)&0.27(3)&6.4(1)\\
$\bf{h_{3}}$&&&$y$&-1.12(6)&-0.32(6)&-0.08(6)&3.6(1)\\
&&&$z$&-0.97(7)&-0.32(6)&0.55(6)&3.4(1)\\
\\-0.15&0.42&${\trhalf}$&$x$&-0.88(10)&-0.11(8)&0.16(8)&2.3(1)\\
$\bf{h_{4}}$&&&-$x$&1.01(4)&0.06(4)&-0.16(4)&3.3(1)\\
&&&$y$&0.44(5)&-0.92(6)&-0.47(5)&3.9(1)\\
&&&$z$&0.56(5)&-0.40(4)&0.76(5)&3.8(4)\\
&&&-$z$&0.24(6)&0.32(5)&-0.88(5)&3.8(3)\\
\\-0.15&0.42&-${\trhalf}$&$x$&-0.93(10)&0.03(8)&0.11(8)&2.0(3)\\
$\bf{h_{5}}$&&&$y$&0.52(6)&-0.73(6)&0.32(5)&3.0(4)\\
&&&$z$&0.54(6)&0.34(5)&0.88(6)&4.2(3)\\
\\0.15&-0.42&${\trhalf}$&$x$&-0.92(6)&-0.10(6)&-0.27(4)&3.3(6)\\
$\bf{h_{6}}$&&&$y$&0.53(4)&-0.84(4)&-0.38(3)&4.4(1)\\
&&&$z$&0.43(4)&-0.61(3)&0.86(3)&3.4(2)\\
\end{tabular}
\end{ruledtabular}
\end{table}
Following the description of a magnetic
structure given in reference,\cite{Zaharko04} we express the moment
${\bf{S}_{jl}}$ of the
$j$th Cu$^{2+}$ ion in the $l$th unit cell as
\begin{equation}
{ {\bf{S}_{jl}} = {\bf{A}_{j}} \cos({\bf{k'\cdot
r_{l}}}+\psi_{j})+{\bf{B}_{j}} \sin({\bf{k'\cdot r_{l}}}+\psi_{j}) }
\label{eq:eq4}
\end{equation}
with ${\bf{r_{l}}}$ being the vector defining the origin of the
$l$th unit cell. ${\bf A_j}$ and ${\bf B_j}$ are orthogonal vectors
which determine
the magnitude and direction of the helix associated with the $j$th ion,
whilst $\psi_{j}$ defines its phase. The 4 independent Cu$^{2+}$
moments of the ${\CuTeCl}$ unit cell
could rotate on independent
helices in which case it would be neccessary to define the plane of
each helix in polar
coordinates by the angles $\theta_j$, $\phi_j$ of ${\bf{B}_{j}}$.
There is freedom to choose the origin of each helix and a convenient
choice is with the vector ${\bf{A}_{j}}$ in the $ab$ plane
($\theta_{\bf{A}_{j}}$=90\deg). For the class of models in which the four
Cu$^{2+}$
ions rotate as two canted pairs\cite{Zaharko04} there are only two planes to
define since the moments on the two ions of a pair rotate in the same
one  ($\theta$, $\phi$ are the same).
The difference between the $\psi$ values  of the two ions is the canting angle
for the pair, $\alpha$.
\\
Least-squares refinement of the $\theta$, $\phi$ and $\psi$
parameters against the polarimetric measurements  for the
reflections $\bf{h_{1}}$-$\bf{h_{6}}$ made using a CCSL program\cite{CCSL}
lead to the following
conclusions:
\begin{itemize}
\item[1.]
The data are sensitive to the difference between the $\phi$ angles of the two
helices which defines the angle between the two planes, and to the
absolute value of $\theta_{\bf{B}}$ which defines their inclination
to the $c$-axis. However, the sensitivity to the absolute
values of the angle $\phi$ which defines their inclination
to the $a$-axis and the phase $\psi$ is not very high.
\item[2.]
The assumption that the
envelope of the helices is circular (${|\bf{A}_{j}|}={|\bf{B}_{j}|}$)
and that all the Cu$^{2+}$ ions have the same moment is
supported by the polarimetric  data. No significant improvement in the fit was
obtained by allowing any of the components of moment to vary.
\item[3.]
The best agreement (Table~\ref{tab2}) was achieved for
a model comprising two pairs of spins with the ${\bf A}$ vectors lying in the
$ab$ plane ($\theta_{\bf{A}}$=90\deg) and
the ${\bf B}$ vectors directed along the $c$ axis
($\theta_{\bf{B}}$=0\deg). The
angle between the two planes on which the spin pairs rotate is small, not
exceeding 10\deg. Allowing the 4  helices to be independent
did not improve the fit.
\item[4.]
To fix other details of the
magnetic structure we need to complement the polarimetric data with the
integrated intensity measurements.
\end{itemize}

\begin{figure}[tbh]
\caption {The $ac$ (top) and $ab$ (bottom) view on the layer of spin 
tetrahedra of the ${\CuTeCl}$ ${\bf{k'}}$ magnetic structure. The 
origin is shifted by [0 0 1/2] relative to the crystallographic unit 
cell.}
\includegraphics[width=86mm,keepaspectratio=true]{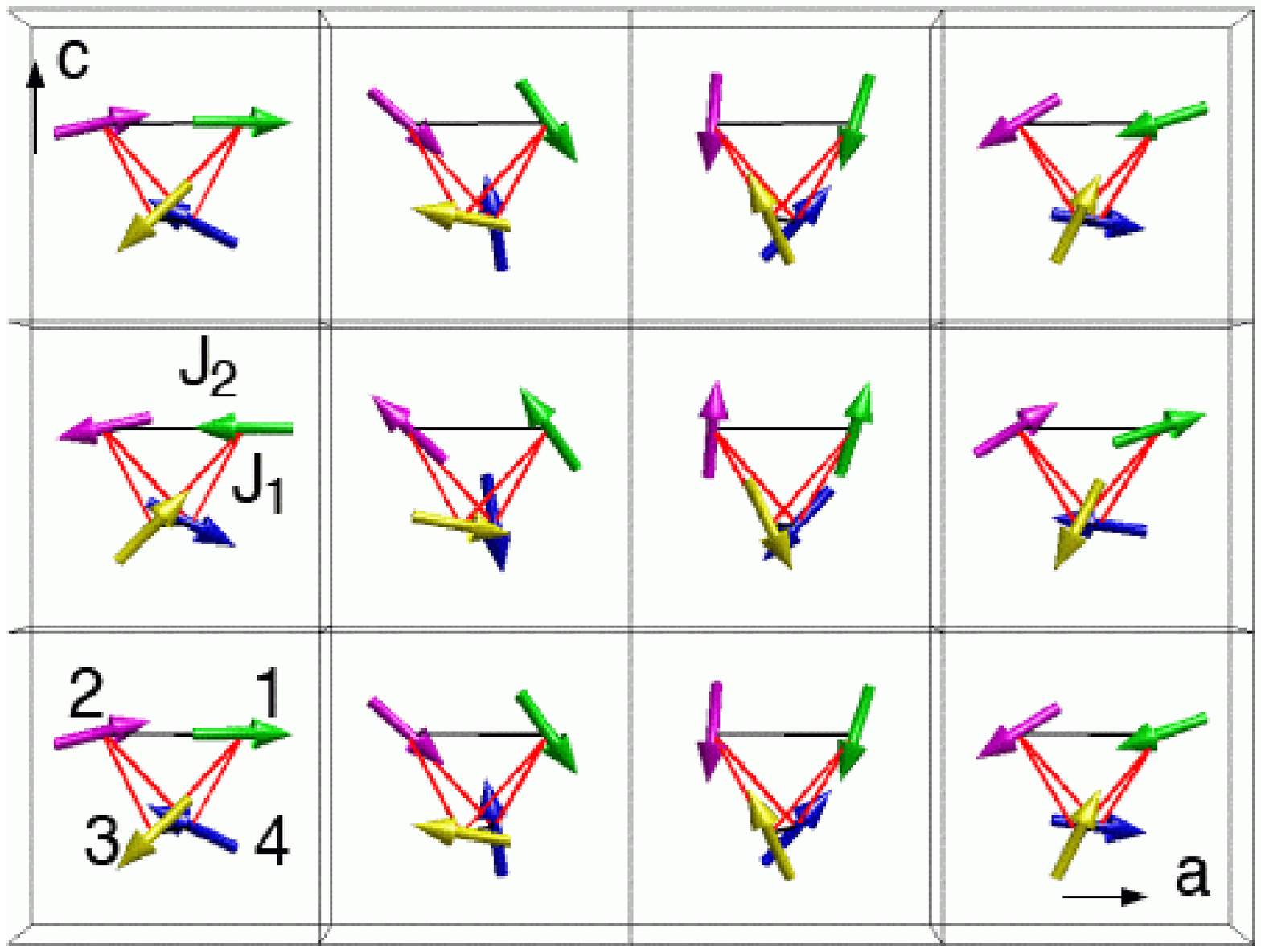}
\includegraphics[width=86mm,keepaspectratio=true]{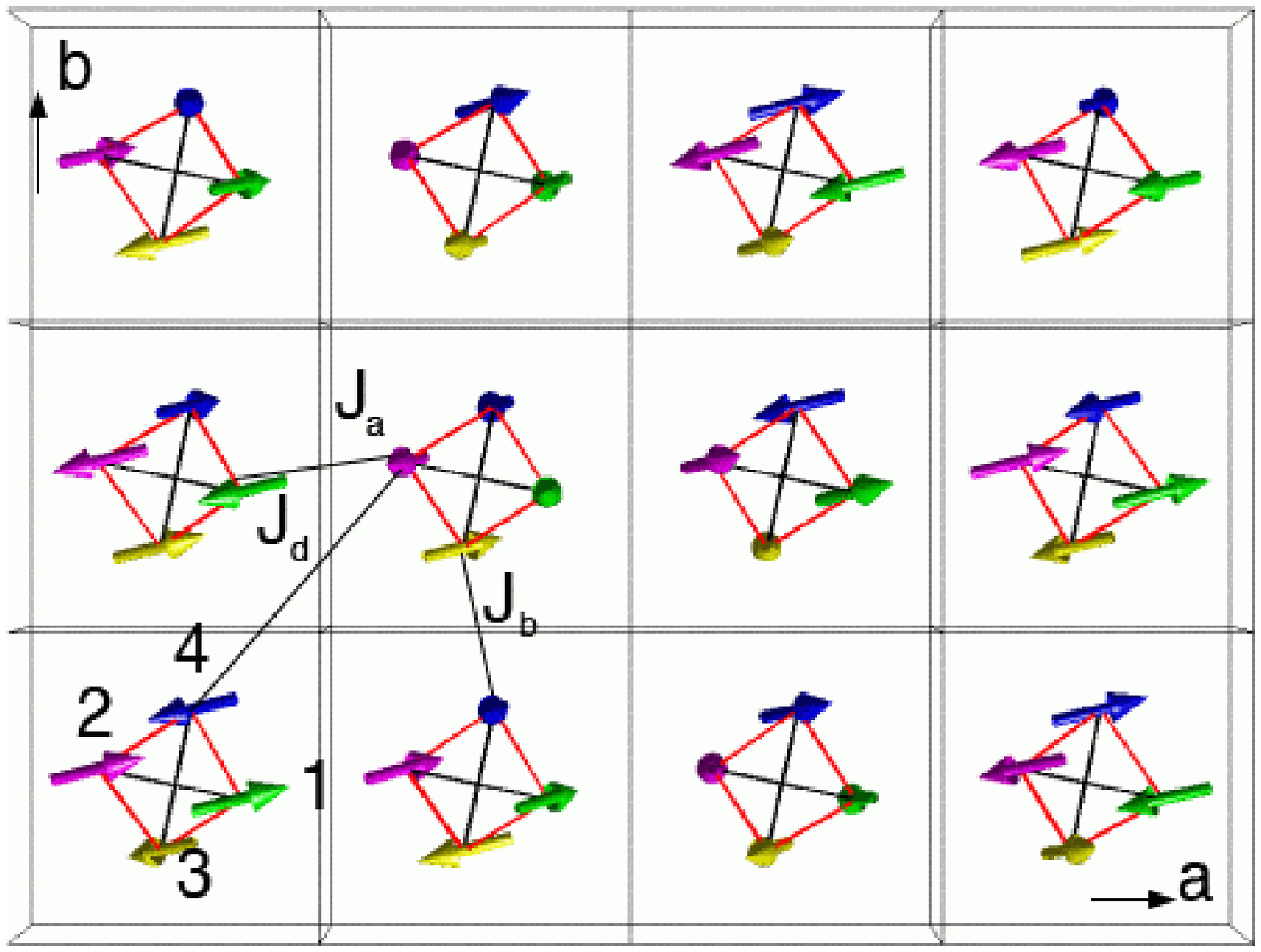}
\label{fig2}
\end{figure}

\begin{figure}[tbh]
\caption {The  $ac$ layer of spin tetrahedra of the ${\CuTeBr}$ ${\bf{k'}}$
magnetic
structure.}
\includegraphics[width=86mm,keepaspectratio=true]{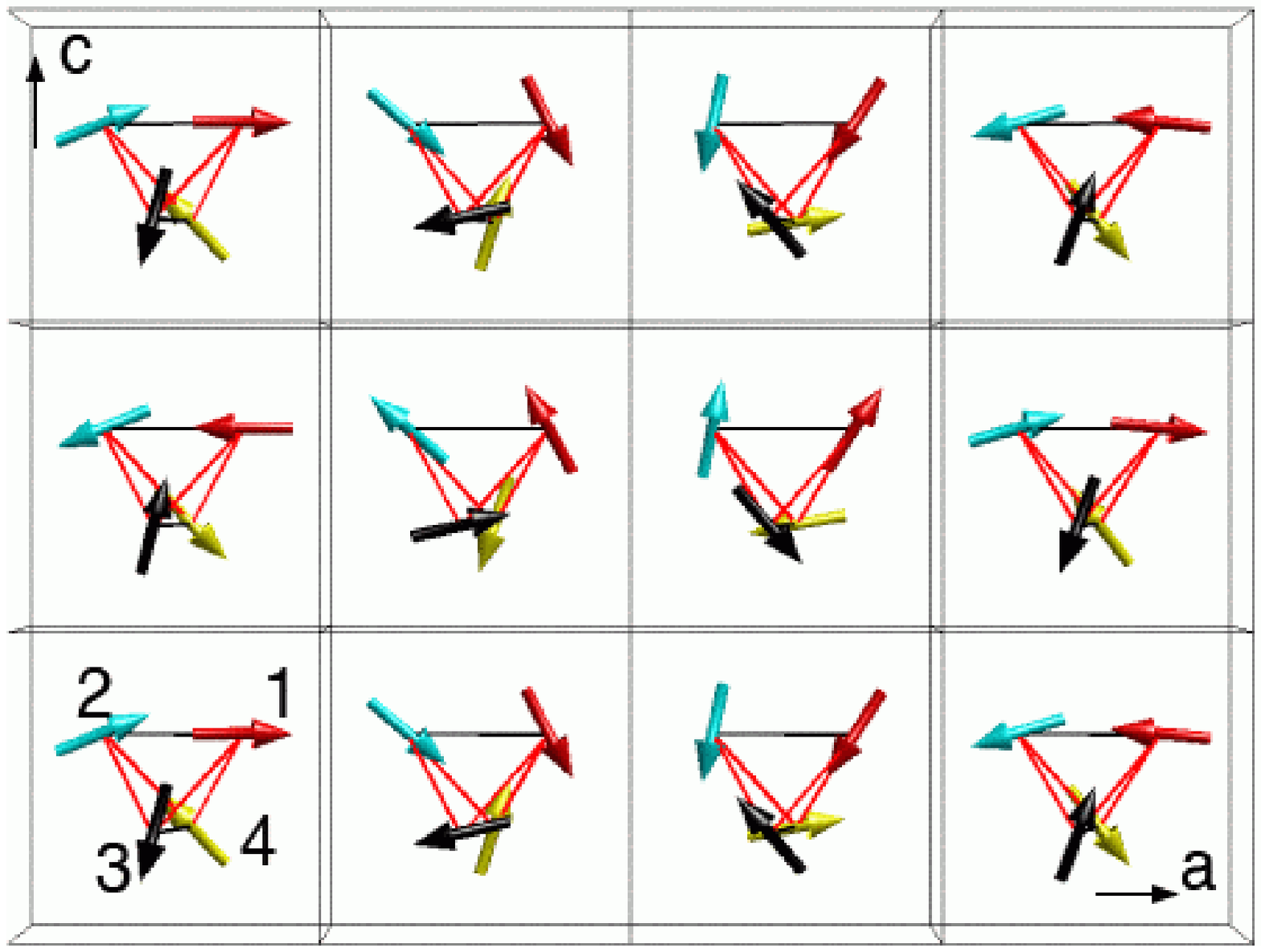}
\label{fig3}
\end{figure}

\begin{table}
\caption{Polarization matrices $P_{ij}$ calculated for final model of the
${\bf{k'}}$ magnetic structure of ${\CuTeCl}$ presented in Table~\ref{tab3}.
\label{tab2}}
\begin{ruledtabular}
\begin{tabular}{ccccccc}
\multicolumn{3}{c}{${h~~~~k~~~~l}$}&${P'_{i}}$&${P_{ix}}$&${P_{iy}}$&${P_{iz}}$\\
\\
0.15&-0.42&${\half}$&-$x$&0.97&0.00&0.26\\
$\bf{h_{1}}$&&&$y$&0.83&-0.53&0.15\\
&&&$z$&0.64&-0.02&0.77\\
&&&-$z$&0.97&0.02&-0.26\\
\\0.15&-0.42&-${\half}$&-$x$&0.97&0.00&0.26\\
$\bf{h_{2}}$&&&$y$&0.83&-0.53&0.19\\
&&&$z$&0.64&0.03&0.77\\
&&&-$z$&0.97&-0.02&-0.26\\
\\-0.15&-0.58&${\half}$&$x$&-0.94&-0.30&-0.13\\
$\bf{h_{3}}$&&&$y$&-0.98&-0.03&0.21\\
&&&$z$&-0.79&-0.34&0.50\\
\\-0.15&0.42&${\trhalf}$&$x$&-0.98&-0.06&0.19\\
$\bf{h_{4}}$&&&-$x$&0.99&0.03&-0.13\\
&&&$y$&0.38&-0.80&-0.47\\
&&&$z$&0.54&-0.42&0.73\\
&&&-$z$&0.28&0.45&-0.85\\
\\-0.15&0.42&-${\trhalf}$&$x$&-0.98&0.06&0.19\\
$\bf{h_{5}}$&&&$y$&0.45&-0.80&0.40\\
&&&$z$&0.54&0.42&0.73\\
\\0.15&-0.42&${\trhalf}$&$x$&-0.98&0.06&-0.19\\
$\bf{h_{6}}$&&&$y$&0.45&-0.80&-0.40\\
&&&$z$&0.28&-0.45&0.85\\
\end{tabular}
\end{ruledtabular}
\end{table}
The unpolarized integrated intensity sets consist of 98 ${\bf{k'}}$
magnetic (and 286 nuclear) reflections for the ${\CuTeCl}$ crystal and 44
magnetic (30 nuclear) for the
${\CuTeBr}$ crystal. Due to the small size of the ${\CuTeBr}$ crystal
and its low magnetic moment, very long counting times were needed; measurement
of each magnetic reflection lasted up to 4.5 h.
Nuclear intensities were corrected for absorption and extinction, which for the
${\CuTeCl}$ crystal was significant.
When modeling the magnetic structure the scale refined from the nuclear
reflections and the parameters reliably determined from the polarimetry
experiment were fixed.
We assumed a constant moment model and constrained ${\bf A}$ to lie in the $ab$
plane and ${\bf B}$ to be parallel to the $c$ axis. The intensity data are
sensitive to the absolute values of the magnetic
moment and the angles $\phi_A$ and $\psi$, in contrast to the
polarimetric data,
and the values obtained from these refinements are
given in Table~\ref{tab3}. The model itself is illustrated schematically in
Fig.~\ref{fig2}. The goodness of fit of the model in which there was a
difference in the
$\phi_A$ angles of two pairs of 10\deg\ was not significantly different from
that in which it was zero,
so within the statistical accuracy all spins
rotate in the same plane.
\\
For ${\CuTeBr}$ the same constraints were used in the refinement of the
integrated intensity data.
The final values are listed in Table~\ref{tab3} and the structure is presented
in Fig.~\ref{fig3}. The refinement was much more stable than for the
${\CuTeCl}$
intensity data and always converged to these final values, even when releasing
the constraints and starting with different initial values. In fact,
a simulated
annealing algorithm~\cite{Rodriguez, Kirkpatrick} was applied to the
generalised
helix model (in which the moments are equal but all other constraints on the
helices are relaxed), and the resulting structure was extremely close to that
presented in Fig.~\ref{fig3}.
\begin{table}
\caption{
The ${\bf{k'}}$ magnetic structure of ${\CuTeX}$ (X=Cl, Br).
The origin of the
helices is chosen in the $ab$ plane ($\theta_A$=90).  The phase of the first
helix $\psi_1$ is set to 0\deg. $\phi_B$=$\phi_A$+90\deg
due to
orthogonality of
${\bf A_j}$ and ${\bf B_j}$.
$\theta_B$ is fixed to zero based on polarization data.
$\alpha_{ij}$ (
$i,j$=1,4) is the canting angle between moments of the Cu$^{2+}$ ions with
coordinates $x\approx 0.730$, $y\approx 0.453$, $z\approx 0.158$: 1 
($x,y,z$), 2 ($1-x,1-y,z$), 3 ($y,1-x,-z$), 4 ($1-y,x,-z$).
\label{tab3}}
\begin{ruledtabular}
\begin{tabular}{ccccccc}

&&m, [$\mu_{B}/Cu$]&$\phi_A$&$\psi_2$&$\psi_3$&$\psi_4$, [\deg]\\
\\
X=Cl&&0.88(1)&14(5)&13(3)&44(3)&-26(4)\\
X=Br&&0.395(5)&9(5)&22(4)&75(5)&-46(3)\\
\hline
&$\alpha_{12}$&$\alpha_{34}$&$\alpha_{13}$&$\alpha_{14}$&$\alpha_{23}$&$\alpha_{24}$, 

[\deg]\\
\\
X=Cl&13&70&135&154&147&142\\
X=Br&22&120&105&134&127&112\\
\end{tabular}
\end{ruledtabular}
\end{table}
\begin{table}
\caption{
Comparison between selected observed and calculated magnetic
structure factors of ${\bf{k'}}$ (present D10 experiment, $F^{\bf{k'}}, h^{\bf{k'}}$) and ${\bf{k}}$ (previous D15 experiment, $F^{\bf{k}}, h^{\bf{k}}$)
reflections of ${\CuTeCl}$ crystals.
\label{tab4}}
\begin{ruledtabular}
\begin{tabular}{ccccccccc}
$h^{\bf{k'}}$&$k$&$l$&$F^{\bf{k'}}_{obs}$&$F^{\bf{k'}}_{calc}$&&$h^{\bf{k}}$&$F^{\bf{k}}_{obs}$&$F^{\bf{k}}_{calc}$\\

\\
    -0.15&0.42&0.5&7.280&6.487&&0.15&8.8130&8.2691\\
      0.15&-0.42&0.5&7.550&6.206&&-0.15&8.4691&8.2893\\
     -0.15&-0.58&0.5& 8.544&7.545&&0.15&7.1362&6.8697\\
     -0.15&-0.58&-0.5&8.602&6.892&&0.15&7.2834&6.8684\\
     -0.85&0.58&0.5&8.000&7.426&&0.85&2.0599&1.1244\\
     -1.15&0.42&-0.5&5.099&5.118&&1.15&2.8397&4.2169\\
     -1.15&-0.58&0.5&5.385&4.851&&1.15&3.2573&3.9668\\
     -0.85&-0.42&0.5& 3.606&2.003&&0.85&5.8267&6.0495\\
     -1.85&-0.42&0.5&5.477&4.607&&1.85&7.1667&6.4599\\
      0.15&-0.42&-1.5&8.660&8.310&&-0.15&8.8371&7.8371\\
      0.15&-0.42&1.5&8.660&8.285&&-0.15&8.5935&7.8208\\
     -0.15&0.42&1.5& 8.718&8.310&&0.15&8.8130&7.8371\\
      0.15&0.58&-1.5&11.747&12.104&&-0.15&5.8990&5.1342\\
     -0.15&-0.58&-1.5&11.874&12.058&&0.15&5.9349&5.1453\\
     -0.15&-0.58&1.5&11.662&12.104&&0.15&6.4821&5.1342\\
     -0.85&0.58&-1.5&6.000&6.474&&0.85&2.0599&0.7765\\
   \end{tabular}
\end{ruledtabular}
\end{table}\\
\\
The model for the ${\bf{k'}}$ structure of ${\CuTeCl}$ developed here gives a
good fit to the limited ${\bf{k'}}$ set of reflections measured
previously.\cite{Zaharko04} This model gives very poor agreement with the
${\bf{k}}$ reflections,  but
significant improvement can be achieved by allowing the planes in which the two
pairs rotate to be inclined to one another in accordance with the previously
determined ${\bf{k}}$ model. An interesting detail is that the
canting angles $\alpha_{12}$ and $\alpha_{34}$ are almost the
same in the ${\bf{k'}}$ X=Cl, Br and ${\bf{k}}$ X=Cl structures of ${\CuTeX}$.

\section{Discussion}

The findings of our experiment, namely, the coexistence, in some crystals,
of two symmetrically independent wave vectors, ${\bf{k'}}$ and ${\bf{k}}$;  two
different magnetic structures, one associated with each wave vector;  two
different configurations for the spins in the Cu$^{2+}$ tetrahedra: the
'canted coplanar' and 'canted pair' motifs in these magnetic
structures, are very
puzzling.\\
The ground state of an isolated tetrahedron with AF
exchange interactions between S=1/2 spins at the vertices is a singlet:
$\sum_{i=1}^4 S_i$=0. No long range magnetic order would exist in a
structure built from such isolated tetrahedra at any temperature. If the
tetrahedra have tetragonal rather than cubic symmetry, as in the present
case, there are two different intra-tetrahedral exchange constants:
nearest neighbour $J_{1}$ and next nearest neighbour $J_{2}$. If $J_{1} >  J_{2}$ the singlet state involves all four spins whereas if $J_{1} < J_{2}$ the spins form two dimers, each dimer individually forming a spin
singlet.\cite{brenig} In the ${\CuTeX}$ system due to strong 
inter-tetrahedral coupling the tetramers and dimers are not true 
singlets and the ground state is magnetically ordered.
\\
The system is
very complex and the ground state spin arrangement is determined by
competition between the geometrically frustrated intra-tetrahedral
coupling, the exchange between tetrahedra and the antisymmetric
Dzyaloshinski-Moriya interactions.
It is possible that the interplay between
these various couplings could result in several different but nearly degenerate
spin configurations. In this case the spin system could be prompted
to adopt one out of several possible
arrangements by perturbations due to
oxygen or copper defects associated with
slight chemical inhomogeneity. This would explain why the coexistence
of ${\bf{k'}}$ and ${\bf{k}}$ is strongly sample dependent. If we
consider the lattice defined by the centers of the tetrahedra, 
ignoring their symmetry, another observation, the equality in the 
lengths of the components
$k_x$ and $k_y$ for the ${\bf{k'}}$ and ${\bf{k}}$ wave vectors,
becomes clear. Such a lattice has full tetragonal symmetry and the ${\bf{k'}}$
and ${\bf{k}}$ wave vectors are symmetrically equivalent. This could
mean that the length of the wave vector is determined by the
inter-tetrahedral exchange and until there is intra-tetrahedral
ordering, the two wave vectors are degenerate.
We suggest that the final
arrangement adopted by the tetrahedra may be determined either by
chance nucleation and growth of one rather than the other 
wave vector or by small alterations in the relative strengths of 
intra-tetrahedral
interactions caused  by crystal inhomogeneities.
  \\
The ${\bf{k'}}$ structure is the one which occurs most frequently in 
the ${\CuTeX}$ crystals  studied up to now by neutron diffraction. 
Its main feature is that the helices of all
spins rotate almost in  a single plane, which is close to (010). 
The 4 spins of each Cu$^{2+}$ tetrahedron form a  canted coplanar motif
which rotates on a single helix with propagation vector
${\bf{k'}}$. The refined moment is
0.88(1)$\mu_{B}$/Cu (X=Cl) and 0.395(5)$\mu_{B}$/Cu (X=Br).\\
The angles between the spins on the sites 1-2 and 3-4 are very
different from  one another: the Cu1-Cu2 spins are almost collinear
with $\alpha_{12}$=13(3)\deg (X=Cl) and $\alpha_{12}$=22(4)\deg 
(X=Br),
while the Cu3-Cu4 arrangement is almost orthogonal 
$\alpha_{34}$=70(4)\deg\  (X=Cl) and
$\alpha_{34}$=120(5)\deg\ (X=Br). 
Noting that the 
overlap between magnetic orbitals associated with the $J_{2}$ path is 
almost zero\cite{valenti, whangbo} this might indicate that the 
intra-tetrahedral $J_{2}$ coupling is rather weak.
On the other hand the  angles between
spins of different pairs in the same tetrahedron differ only slightly 
(see Table~\ref{tab3})
and are close to 145\deg\ (X=Cl) and 120\deg\ (X=Br). These angles 
are the same for all tetrahedra in the structure and such regularity 
might imply that the $J_{1}$ coupling mediated through the Cu-O-Cu 
superexchange path ($\angle$Cu-O-Cu$\approx$~110\deg) is strong. \\ 

Analysis of the angles between spins in adjacent tetradedra reveals 
that neigbouring ions across the $[1\pm1\ 0]$ diagonals are almost antiparallel
(Fig.~\ref{fig2} bottom).
This implies that the 
inter-tetrahedral diagonal $J_{d}$ coupling could be important with the 
linear superexchange path Cu-X..X-Cu  providing a strong AF 
interaction.\cite{valenti,whangbo}
The angles between neighbouring ions ions related by the [100] and 
[010] lattice translations are very different, in spite of the 
underlying tetragonal symmetry. One, for [100], is acute ($\alpha\approx40$\deg) 
and the other obtuse ($\approx$140\deg~X=Cl, $\approx$110\deg~X=Br)
implying weak $J_{a}$ and $J_{b}$ coupling in accord with 
band-structure calculations.\cite{valenti}\\
\begin{figure}[tbh]
\caption {The $ac$ layer of resultant moment of tetrahedra of the
${\CuTeCl}$ ${\bf{k'}}$ magnetic structure.}
\includegraphics[width=86mm,keepaspectratio=true]{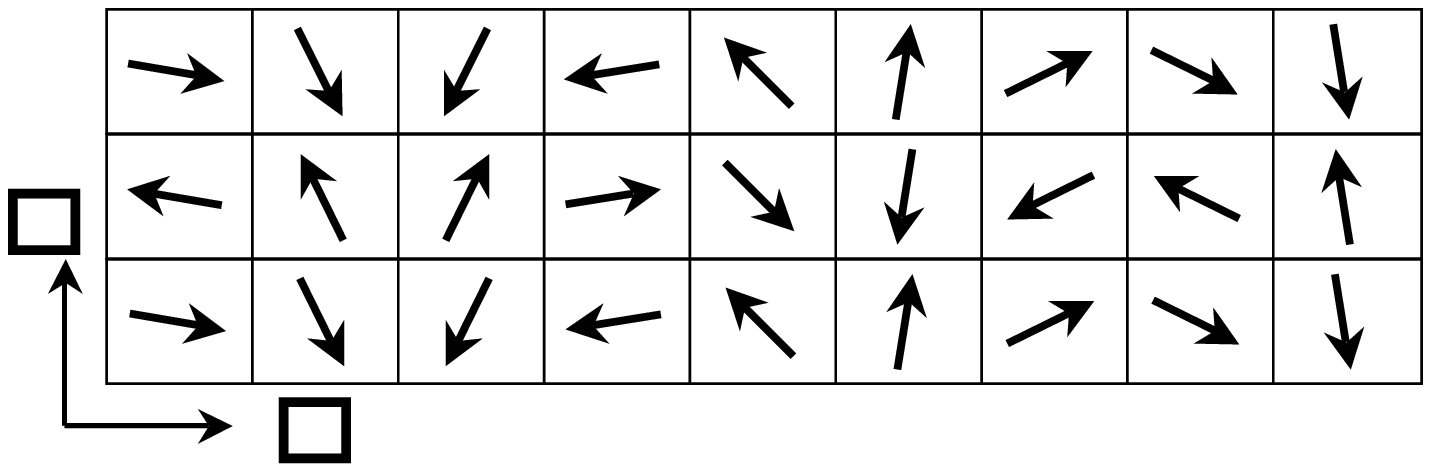}
\label{fig4}
\end{figure}
One further property of the 
proposed ${\bf{k'}}$ spin arrangement should be discussed. 
This model leads to a finite resultant moment on each tetrahedron
which is constant throughout the whole crystal. This moment
rotates in the $ac$ plane (Fig.~\ref{fig4}) on a cycloid with the 
propagation vector ($-k_x$, 0, $\half$) giving an angle  of  
54\deg\ (X=Cl) and 62\deg\ (X=Br) between the neighbouring tetrahedra along $a$. 
Interestingly, the resultant moments on the tetrahedra in 
the chloride (0.333 $\mu_{B}$) and bromide (0.388 $\mu_{B}$) are almost 
equal, although the moment of the Cu$^{2+}$ ions is close to the 
saturated value of 1 $\mu_{B}$/Cu for  ${\CuTeCl}$, whereas it is 
significantly less for ${\CuTeBr}$.\\

 As the S=1/2 Cu$^{2+}$ ion has very little single ion anisotropy, it is not 
clear what is responsible for the choice of the $ac$ plane as the  
easy plane of the spins. It might be either the 
anisotropy of the inter-tetrahedral interactions or the DM 
interactions the direction of which is determined 
by the symmetry of the local environment.\cite{moria}
The DM interaction could be nonzero in the ${\CuTeX}$ system 
and would give a DM vector  in the $ab$ plane\cite{mila05} perpendicular to each Cu-Cu bond
within the tetrahedra. This
antisymmetric coupling would favor two spins to cant in opposite directions in the plane perpendicular to the DM vector and the fairly constant angle between nearest neighbour spins could reflect the ratio DM/$J_1$.
\\
A thorough theoretical study is needed to clarify a number of
questions raised by our findings.
\begin{itemize}
\item[1.] What relative strengths of the $J_1$, $J_2$ intra-tetrahedral
and $J_c$, $J_d$, $J_x$\cite{remark2} inter-tetrahedral couplings
are needed to give
the experimentally observed ${\bf{k'}}$ and ${\bf{k}}$ structures.
\item[2.] Can anisotropy of the inter-tetrahedral interactions alone
explain the easy plane of the magnetic moments in the ${\bf{k'}}$ structure.
\item[3.] Does the choice of wave vector
(${\bf{k'}}$ or ${\bf{k}}$) determine the final spin arrangement
adopted by the tetrahedra or do changes in strength of the $J_1$,
$J_2$ couplings moderate the choice between the
${\bf{k'}}$ (`canted coplanar' ) and the  ${\bf{k}}$ (`canted pair') 
structures.
\end{itemize}

\acknowledgments
This work was carried out  at the ILL reactor, Grenoble,
France and on SINQ at the Paul Scherrer Institute, Villigen, Switzerland.
We would like to thank Drs. N. Kernavanois and M. Medarde for
experimental assistance during the D3 experiment and Dr. A. M. Mulders 
for help during the X10 experiment. The sample preparation was
supported by the
NCCR research pool MaNEP of the Swiss NSF. SJC acknowledges the
financial support of the UK Engineering and Physical Sciences
Research Council.


\begin{thebibliography}{99}
\bibitem{remark2}
Notations from Ref.\onlinecite{valenti} are adopted.
\end{thebibliography}
\end{document}